\documentclass[aip, amsmath, amssymb, reprint, floatfix]{revtex4-1}% 

\usepackage[version=3]{mhchem}
\usepackage{graphicx}
\usepackage{bm}
\DeclareMathOperator*{\argmin}{arg\, min}

\begin{document}

\title{CLEASE: A versatile and user-friendly implementation of Cluster Expansion method}

\author{Jin Hyun Chang}
\email{jchang@dtu.dk}
\affiliation{Department of Energy Conversion and Storage, Technical University of Denmark, DK-2800 Kgs. Lyngby, Denmark}

\author{David Kleiven}
\affiliation{Department of Physics, Norwegian University of Science and Technology, NO-7491 Trondheim, Norway}

\author{Marko Melander}
\affiliation{Department of Chemistry, University of Jyv\"{a}skyl\"{a}, Jyv\"{a}skyl\"{a}, Finland}

\author{Jaakko Akola}
\altaffiliation{Laboratory of Physics, Tampere University of Technology, FI-33101 Tampere, Finland}
\affiliation{Computational Physics Laboratory, Tampere University, P.O. Box 692, FI-33014 Tampere, Finland}

\author{Juan Maria Garcia Lastra}
\author{Tejs Vegge}
\affiliation{Department of Energy Conversion and Storage, Technical University of Denmark, DK-2800 Kgs. Lyngby, Denmark}     

\date{\today}

\begin{abstract}
Materials exhibiting a substitutional disorder such as multicomponent alloys and mixed metal oxides/oxyfluorides are of great importance in many scientific and technological sectors. Disordered materials constitute an overwhelmingly large configurational space, which makes it practically impossible to be explored manually using first-principles calculations such as density functional theory (DFT) due to the high computational costs. Consequently, the use of methods such as cluster expansion (CE) is vital in enhancing our understanding of the disordered materials. CE dramatically reduces the computational cost by mapping the first-principles calculation results on to a Hamiltonian which is much faster to evaluate. In this work, we present our implementation of the CE method, which is integrated as a part of the Atomic Simulation Environment (ASE) open-source package. The versatile and user-friendly code automates the complex set up and construction procedure of CE while giving the users the flexibility to tweak the settings and to import their own structures and previous calculation results. Recent advancements such as regularization techniques from machine learning are implemented in the developed code. The code allows the users to construct CE on any bulk lattice structure, which makes it useful for a wide range of applications involving complex materials. We demonstrate the capabilities of our implementation by analyzing the two example materials with varying complexities: a binary metal alloy and a disordered lithium chromium oxyfluoride.
\end{abstract}


\maketitle

\section{Introduction}
Computational modeling of materials with a substitutional disorder such as multicomponent alloys and mixed metal oxides is said to have a \textit{configurational} problem. The vast configurational space of these materials makes it practically impossible to explore directly using first-principles calculations such as density functional theory (DFT). A quantitative method capable of establishing the relationship between the structure and property of materials is therefore essential. Cluster Expansion (CE) \cite{Sanchez1984, DeFontaine1994, Zunger1993, Asta2001, VandeWalle2008, Zhang2016} is a method that has been used successfully in the past few decades to parameterize and express the configurational dependence of physical properties. The most widely parameterized physical property is energy computed using first-principles methods, but CE can also be used to parameterize other quantities such as band gap \cite{Magri1991, Franceschetti1999} and density of states \cite{Geng2005}. 

Despite its success and usefulness in predicting physical properties of crystalline materials, CE remains as a niche tool used in a small subfield within the computational materials science, primarily used by specialists. On the other hand, the research fields in which CE is becoming relevant is on the rise; one such example is the use of disordered materials for battery applications \cite{Wang2015a, Abdellahi2016, Abdellahi2016a, Urban2016, Kitchaev2018}. The objective of our work is to make cluster expansion more accessible for a broad range of computational scientists who do not necessarily possess expertise in cluster expansion. Our approach to achieving such a goal is to implement CE as a part of a widely used, open-source Atomic Simulation Environment (ASE) package \cite{ASE}. Henceforth, we refer to our implementation as CLEASE, which stands for CLuster Expansion in Atomic Simulation Environment. 

Having CE as a part of a widely used package with interfaces to a multitude of open-source and commercial atomic-scale simulation codes accompanies several practical benefits: (1) a large existing user base does not need to install or learn a new program as the CE module is a part of ASE and inherits its syntax and code style, and (2) all of the atomic-scale simulation codes supported by ASE are also automatically supported by the implemented module. In addition, CLEASE utilizes the database management feature implemented in ASE, which provides an efficient way to store, maintain and share both DFT and CE results. Therefore, the implementation presented in this article appeals to a significant portion of computational materials science community as a versatile and easy-to-learn package, thereby lowering the barrier to incorporate cluster expansion as a part of their research methods to accelerate computational materials prediction and design. 

The rest of the paper is organized as follows. A brief overview of cluster expansion formalism and other important concepts are provided in section~\ref{sec:theory} in order to aid the readers who are not familiar with the cluster expansion method. The implementation of CLEASE is described in section~\ref{sec:implementation}. Section~\ref{sec:example} contains two application examples with different levels of complexities, namely a binary metal alloy and a lithium metal oxyfluoride. The computational settings and technical details for the examples are provided in section~\ref{sec:methods}.

\section{Theory}
\label{sec:theory}
\subsection{Cluster Expansion Formalism}
The core concept of the cluster expansion is to express the scalar physical quantity of a material, $q(\bm{\sigma})$, to its configuration, $\bm{\sigma}$, where a crystalline system is represented with a fixed underlying grid of atomic sites. In such a representation, any configuration with the same underlying topology can be completely specified by the atomic occupation of each atomic site. For the case of a crystalline material with $N$ atomic sites, any configuration can be specified by an $N$-dimensional vector $\bm{\sigma} = \{s_1, s_2, \ldots, s_N\}$, where $s_i$ is a site variable that specifies which type of atom occupies the atomic site $i$ (also referred to as an occupation variable \cite{Zarkevich2004, Meng2009, VandeWalle2009} or pseudospin \cite{DeFontaine1994, Magri1991, Nelson2013, Nelson2013a, Seko2014}). It is noted that the terms configuration and structure are often used interchangeably.

For the case of multinary systems consisting of $M$ different atomic species, $s_i$ takes one of $M$ distinct values. The original formulation of Sanchez et al. \cite{Sanchez1984} specifies the $s_i$ to take any values from $\pm m$, $\pm (m-1)$, $\ldots$, $\pm 1$ for $M = 2m$ (for the case where there is an odd number of element types, an additional value of 0 should be included in the possible values of $s_i$, and the relation between $M$ and $m$ becomes $M = 2m-1$). Other choices of $s_i$ are also commonly used such as values ranging from $0$ to $M-1$ by van de Walle \cite{VandeWalle2009} and from $1$ to $M$ by Mueller and Ceder \cite{Mueller2010}. Based on the original formalism by Sanchez et al., single-site basis functions are determined through an orthogonality condition
\begin{equation}
    \frac{1}{M} \sum_{s_i=-m}^{m}\Theta_n(s_i)\Theta_{n'}(s_i) = \delta_{nn'}, 
\label{eq:orthogonality_condition}
\end{equation}
where $\Theta_{n}(s_i)$ is the $n$th single-site basis function (e.g., Chebyshev polynomials) for $i$th site and $\delta_{nn'}$ is a Kronecker delta.

\begin{figure*}[!htb]
	\centering
 	\includegraphics[width=150mm]{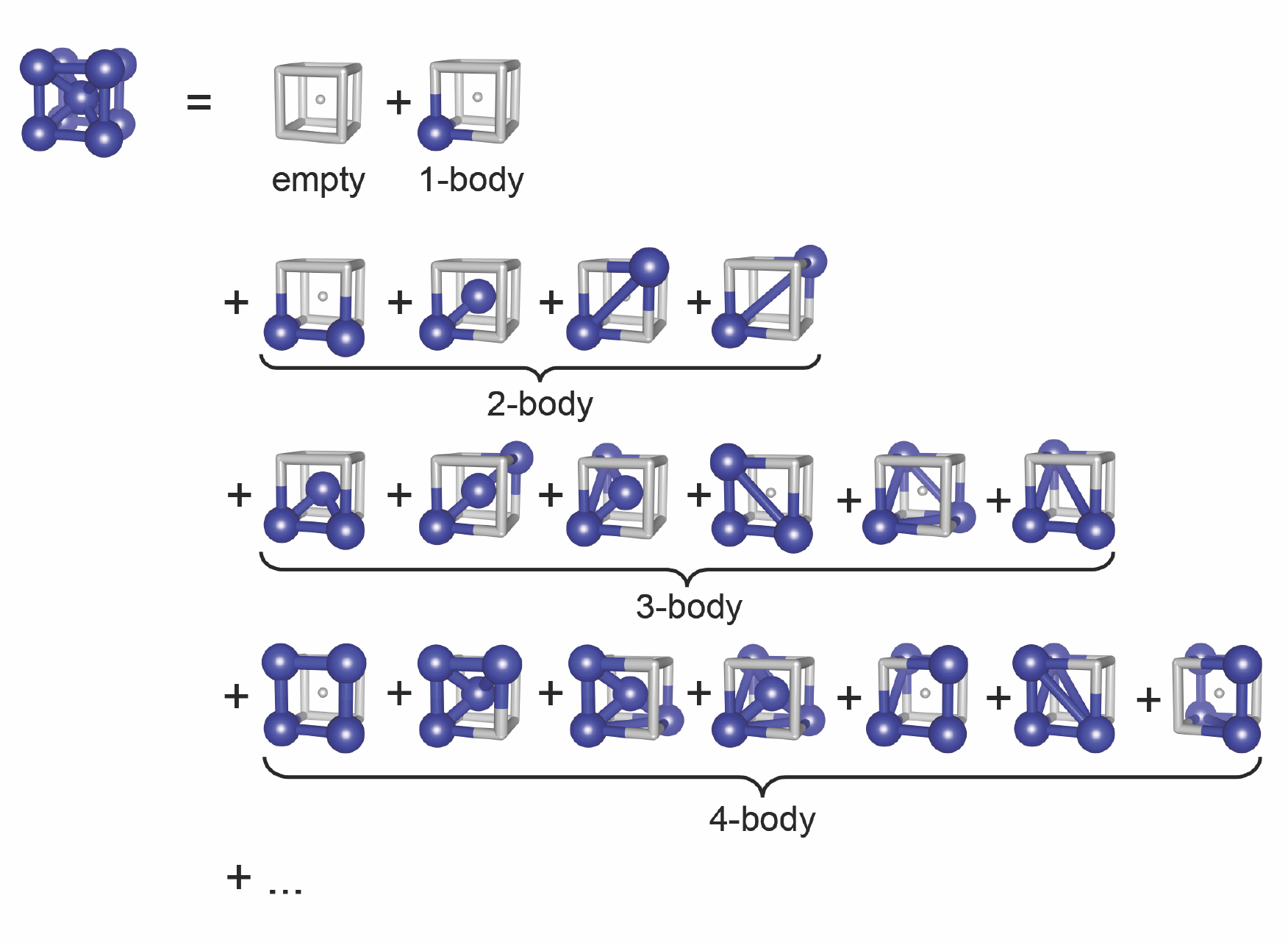}
  	\caption{A simplified illustration of the decomposition of a body-centered cubic lattice.}
\label{fig:cluster_decomposition}
\end{figure*}

The configuration is decomposed into a sum of clusters as shown in figure~\ref{fig:cluster_decomposition}. Each cluster has a set of associated cluster functions, which are defined as 
\begin{equation}
    \Phi_{\bm{n}}(\bm{s}) = \prod_{i} \Theta_{n_i}(s_i),
\label{eq:cluster_function}
\end{equation}
where $\bm{n}$ and $\bm{s}$ are vectors specifying the order of the single-site basis function and the site variables in the cluster, respectively. $n_i$ and $s_i$ specify the $i$th element of the respective vectors. The use of orthogonal basis functions guarantees that the cluster functions defined in (\ref{eq:cluster_function}) are also orthogonal. The symmetrically equivalent clusters are classified as the same cluster, and the collection of all symmetrically equivalent clusters are denoted with an $\alpha$. 

The average value of the cluster functions in cluster $\alpha$ is referred to as a correlation function, $\phi_{\alpha}$. The physical quantity, $q(\bm{\sigma})$, normalized with the number of atomic sites $N$ is then expressed as 
\begin{equation}
    q(\bm{\sigma}) = \sum_{\alpha} m_{\alpha} J_{\alpha} \phi_{\alpha},
\label{eq:cluster_expansion1}
\end{equation}
where $m_{\alpha}$ is the multiplicity factor indicating the number of cluster $\alpha$ per atom and $J_{\alpha}$ is the effective cluster interaction (ECI) per occurrence, which needs to be determined. It is noted that the cluster $\alpha$ includes the cluster of size zero, which have $m_{\alpha} \phi_{\alpha} = 1$. Alternatively, (\ref{eq:cluster_expansion1}) can be written in a more explicitly form,
\begin{equation}
    q(\bm{\sigma}) = J_{0} + \sum_{\alpha} m_{\alpha} J_{\alpha} \phi_{\alpha},
\label{eq:cluster_expansion2}
\end{equation}
where $J_0$ is the ECI of an empty cluster while $\alpha$ in this case corresponds to the clusters of size one and higher. It is often more practical and convenient to express the ECI per atom rather than per occurrence \cite{Zhang2016}, in which case $m_{\alpha}$ and $J_{\alpha}$ are combined into one term, $\widetilde{J}_{\alpha}= m_{\alpha}J_{\alpha}$ and (\ref{eq:cluster_expansion1}) becomes
\begin{equation}
    q(\bm{\sigma}) = \sum_{\alpha} \widetilde{J}_{\alpha} \phi_{\alpha}.
\label{eq:cluster_expansion3}
\end{equation}
CLEASE uses the ECI per atom ($\widetilde{J}_{\alpha}$), but interested users can determine the value of $J_{\alpha}$ based on the values of $m_{\alpha}$ and $\widetilde{J}_{\alpha}$. 
 
Theoretically, there is an infinite number of terms in (\ref{eq:cluster_expansion3}) for an infinite crystal, and the resulting expression can represent any scalar function $q(\bm{\sigma})$ given that appropriate ECI values are found. In practice, sufficient accuracy is often reached with clusters with small number of atoms (e.g., one-, two- and three-body clusters) that are relatively compact in size (e.g., 5 to 7 \AA{} in diameter).

\subsection{Cluster Selection \& Determination of ECI Values}
A crucial element of CE approach is to select relevant clusters from a theoretically infinite number of possible clusters. Many multicomponent systems yield thousands of clusters even when the expansion is limited to relatively compact size and small number of atoms, and they are vastly truncated since only a small fraction of them is needed to achieve the required accuracy. Determining the optimal set of clusters that minimizes the number of clusters without losing its predictive power has been a topic of keen interest in the past decade \cite{Zarkevich2004, Blum2005, Hart2005, Diaz-Ortiz2007, Lerch2009}, and the cluster selection based on genetic algorithms \cite{Blum2005, Hart2005, Lerch2009} was considered to be the most robust method. 

More recently, the use of compressive sensing \cite{Nelson2013a} was proposed to efficiently select the clusters and determine their ECIs in one shot. The compressive sensing is based on $\ell_1$ norm (a special case of $\ell_p$ norm where $p = 1$), which is defined as 
\begin{equation}
    ||\mathbf{x}||_{p} = \left( \sum_{i} \left| x_i \right|^p \right)^{1/p},
\end{equation}
where $\mathbf{x}$ is a vector quantity. It is noted that cluster expansion defined in (\ref{eq:cluster_expansion3}) is in the same form as a linear regression model in statistics and machine learning. Therefore, one can treat CE as a linear regression problem and apply regularization techniques based not only on $\ell_1$ norm but also on any other $p$ values, although $\ell_1$ and $\ell_2$ norms are most commonly used. 

The use of regularization techniques for CE can be illustrated by expressing (\ref{eq:cluster_expansion3}) in a matrix form,
\begin{equation}
    \mathbf{q} = \mathbf{X} \bm{\omega}.
\end{equation}
$\mathbf{X}$ is a matrix containing the correlation functions of the training data where each element in row $i$ and column $\alpha$ is defined as 
\begin{equation}
    \mathbf{X}_{i\alpha} = \phi_{\alpha}(\bm{\sigma}_i).
\end{equation}
$\mathbf{q}$ is a column vector in which the $i$th element is the physical quantity $q$ of the configuration $\bm{\sigma}_{i}$ and $\bm{\omega}$ is a column vector in which $\alpha$th element is $\widetilde{J}_{\alpha}$. 

The simplest way of determining $\bm{\omega}$ is by using ordinary least squares (OLS) method, which minimizes the residual sum of squared errors (RSS). RSS is defined as 
\begin{equation}
    \mathrm{RSS} = ||\mathbf{X} \bm{\omega} - \mathbf{q}||_{2}^{2},
\end{equation}
and the minimization of the RSS has a unique solution $\hat{\bm{\omega}}$ where 
\begin{align}
    \begin{split}
        \hat{\bm{\omega}} &= \argmin_{\bm{\omega}} ||\mathbf{X} \bm{\omega} - \mathbf{q}||_{2}^{2}\\
        &=(\mathbf{X}^T\mathbf{X})^{-1}\mathbf{X}^{T} \mathbf{q}.
    \end{split}
    \label{eq:OLS}
\end{align}
The OLS has two major drawbacks \cite{Nelson2013a}. The first is the requirement on which the number of configurations in the training set needs to be larger than the number of clusters being considered. The matrix $\mathbf{X}^{T}\mathbf{X}$ becomes singular in such a case, and the limitations imposed by the first requirement become more severe for systems consisting of many element types since even strict expansion conditions (i.e., small number of atoms per cluster and compact size) can lead to a large number of clusters. The second drawback is the susceptibility to possible overfitting, which refers to the conditions in which the ECI values are over-tuned to accurately represent $q(\bm{\sigma})$ of the training set at a cost of losing its predictive power for the new configurations that are not included in the training set. The overfitting also makes the model prone to noise present in the training data because the model attempts to meticulously fit the model to the training data including the noise therein.

Regularization is an efficient technique to address the aforementioned drawbacks of OLS by adding a regularization term to (\ref{eq:OLS}). The most common regularization scheme are $\ell_1$ and $\ell_2$ regularization, which respectively uses $\ell_1$ and $\ell_2$ norm as a regularization term. For $\ell_1$ regularization, the solution becomes 
\begin{equation}
    \hat{\bm{\omega}} = \argmin_{\bm{\omega}} ||\mathbf{X} \bm{\omega} - \mathbf{q}||_{2}^{2} + \lambda ||\omega||_1,
\label{eq:l1_regularization}
\end{equation}
where $\lambda$ is a regularization parameter that controls the weight given to the regularization term. The main benefit of $\ell_1$ regularization is its promotion of sparsity. In context of CE, the sparsity means a selection of a fewer number of clusters, or many clusters with their ECI values set to zero. It is noted that there is no unique analytical solution for (\ref{eq:l1_regularization}), and it needs to be solved iteratively. Unlike $\ell_1$ regularization, $\ell_2$ regularization has a unique analytical solution which is expressed as 
\begin{align}
    \begin{split}
        \hat{\bm{\omega}} &= \argmin_{\bm{\omega}} ||\mathbf{X} \bm{\omega} - \mathbf{q}||_{2}^{2} + || \bm{\omega} ||_{2}^{2}\\
        &=(\mathbf{X}^T\mathbf{X} + \lambda \mathbf{I})^{-1}\mathbf{X}^{T} \mathbf{q}.
    \end{split}
    \label{eq:l2_regularization}
\end{align}
However, $\ell_2$ regularization does not promote sparsity, and the resulting solution is likely to contain more clusters than necessary. It is noted that Bayesian compressive sensing  \cite{Nelson2013} scheme is introduced for cluster expansion, which effectively eliminates the parameter $\lambda$ in $\ell_1$ and $\ell_2$ regularization schemes while promoting sparsity. 

Regardless of the fitting technique used, the predictive power of the expansion needs to be assessed to determine its accuracy and reliability. Cross-validation (CV) is a technique used for assessing the prediction accuracy of the model. A leave-one-out (LOO) scheme is most commonly used in CE community, and the LOOCV score is defined as
\begin{equation}
    \mathrm{LOOCV} = \left(\frac{1}{N_\textrm{config}} \sum_{i=1}^{N_\textrm{config}} \left(\hat{q}_{i} - q_i \right)^2\right)^{1/2},
\label{eq:loocv}
\end{equation}
where $N_\textrm{config}$ is the number of configurations in the training set, $\hat{q}_{i}$ is the physical quantity of a structure $i$ predicted by CE using $N_\textrm{config} - 1$ structures without a structure $i$ and $q_i$ is the calculated physical quantity of structure $i$. While OLS has only one (likely overfitted) solution, $\ell_1$ and $\ell_2$ regularization schemes have a solution for each $\lambda$ value. The solution --- a selection of clusters and their ECI values --- that yields the lowest LOOCV score is chosen. Although LOO is the most common cross validation scheme in cluster expansion community, $k$-fold CV is one of the most common schemes used in machine learning community. In a $k$-fold CV scheme, the pool of configurations are randomly partitioned into $k$ parts of equal size. The structures in $k-1$ parts are used as training data while the remaining one part is used as a validation set, and the cross validation is repeated $k$ times. 

\subsection{Thermodynamics in Lattice Models}
The true benefit of CE is in its ability to predict the expanded scalar quantity $q(\bm{\sigma})$ based on trained data. An accurate prediction can be made if the CV score of the expanded $q(\bm{\sigma})$ is sufficiently low, and the prediction speed is very fast on modern computer architecture since it only involves executions of only a small number of simple numerical calculations specified in (\ref{eq:cluster_expansion3}). Such a speed boost allows one to conduct types of analyses that require substantial statistical sampling.

In contrast to zero temperature studies where the system occupies the state with lowest energy, an ensemble of configurations with the lowest free energy are occupied at finite temperature. The free energy $G$ is given by \cite{andersen2012introduction} 
\begin{equation}
    G = -\frac{\ln Z}{\beta}, 
    \label{eq:free_energy}
\end{equation}
where $\beta = 1/k_B T$ and $Z$ is the partition function. $k_B$ is the Boltzmann constant and $T$ is temperature in Kelvin. It is noted that the DFT energies are obtained for fully relaxed structures without any external forces or pressure. Thus, the resulting thermodynamic quantities are effectively obtained in the NPT ensemble (fixed number of particles, fixed pressure and fixed temperature). However, the energy predicted by CE is only valid for the volume leading to the minimum energy of a particular atomic arrangement, and the volume fluctuations are neglected. The free energy can be calculated by utilizing the exact differential
\begin{align}
    \begin{split}
    \mathrm{d}(\beta G) &= -\frac{\partial \ln Z}{\partial \beta} \mathrm{d}\beta \\
    &= U \mathrm{d}\beta
    \end{split}
    \label{eq:dF_canonical}
\end{align}
where $U$ is the average internal energy. The free energy can be obtained by a thermodynamic integration from a reference temperature $\beta_\mathrm{ref}$ where $G$ is known, which can be written as \cite{Tuckerman2010}
\begin{equation}
    \beta G = (\beta G)_\mathrm{ref} + \int_{\beta_\mathrm{ref}}^{\beta} \mathrm{d}\beta ' U(\beta').
    \label{eq:G_int_C}
\end{equation}
Important information of the materials under study such as the stability of ordered/disordered  phases can be determined by comparing the free energy of the material at a given composition with respect to the free energies in the pure phases of its constituents.

\section{Implementation}
\label{sec:implementation}
CLEASE utilizes the existing classes and methods of ASE to perform necessary manipulations and analyses for carrying out CE. Among many adopted features, the most noteworthy are the use of 
\begin{itemize}
    \item an \texttt{Atoms} object to represent an atomic configuration ($\bm{\sigma}$),
    \item a built-in database to efficiently store, maintain and share settings, atomic configurations of the training set, values of the correlation functions ($\phi_{\alpha}(\bm{\sigma})$) and DFT energies,
    \item Python programming language and modular design to remove the strict input file/format requirements and to enable easy implementation of new features, and
    \item a \texttt{Calculator} class to determine the physical quantity $q(\bm{\sigma})$ of a new configuration based on its correlation functions and their ECI values.
\end{itemize}
It is noted that the evaluation of correlation functions of a new configuration and the determination of physical quantity, $q(\bm{\sigma})$, based on ECI values can be a slow process using Python programming language. It is especially true for carrying out Monte Carlo simulations after the CE model training is complete. CLEASE includes an optional external module written in C++ programming language that can be installed to accelerate the critical and repetitive calculations, but the usage of the code remains unchanged even when the external module is installed (i.e., CLEASE automatically determines if the C++ add-on is installed, and uses the C++ version if it is present). 

The inheritance of the existing features of ASE allows CLEASE to be fully integrated to ASE where the users can incorporate CE as a part of their research without losing the continuity with the rest of their workflow. The existing users of ASE do not have to install or learn a new CE program nor select a particular DFT package that a CE code supports. In addition to the benefits of integrating CE as a part of ASE, highlights of the features that makes CLEASE versatile and user-friendly include:
\begin{itemize}
    \item a multicomponent cluster expansion that goes beyond binary systems,
    \item a support for several types of basis functions (e.g., basis functions by Sanchez et al. \cite{Sanchez1984}, Van de Walle \cite{VandeWalle2009} and Zhang and Sluiter \cite{Zhang2016}) for a comparison and compatibility with other CE codes,
    \item many methods for selecting clusters and determining ECI values such as OLS, $\ell_1$ and $\ell_2$ regularization schemes, Bayesian compressive sensing and genetic algorithm, and
    \item both leave-one-out and $k$-fold cross validation schemes.
\end{itemize}

\begin{figure}[!htb]
	\centering
	\includegraphics[width=86mm]{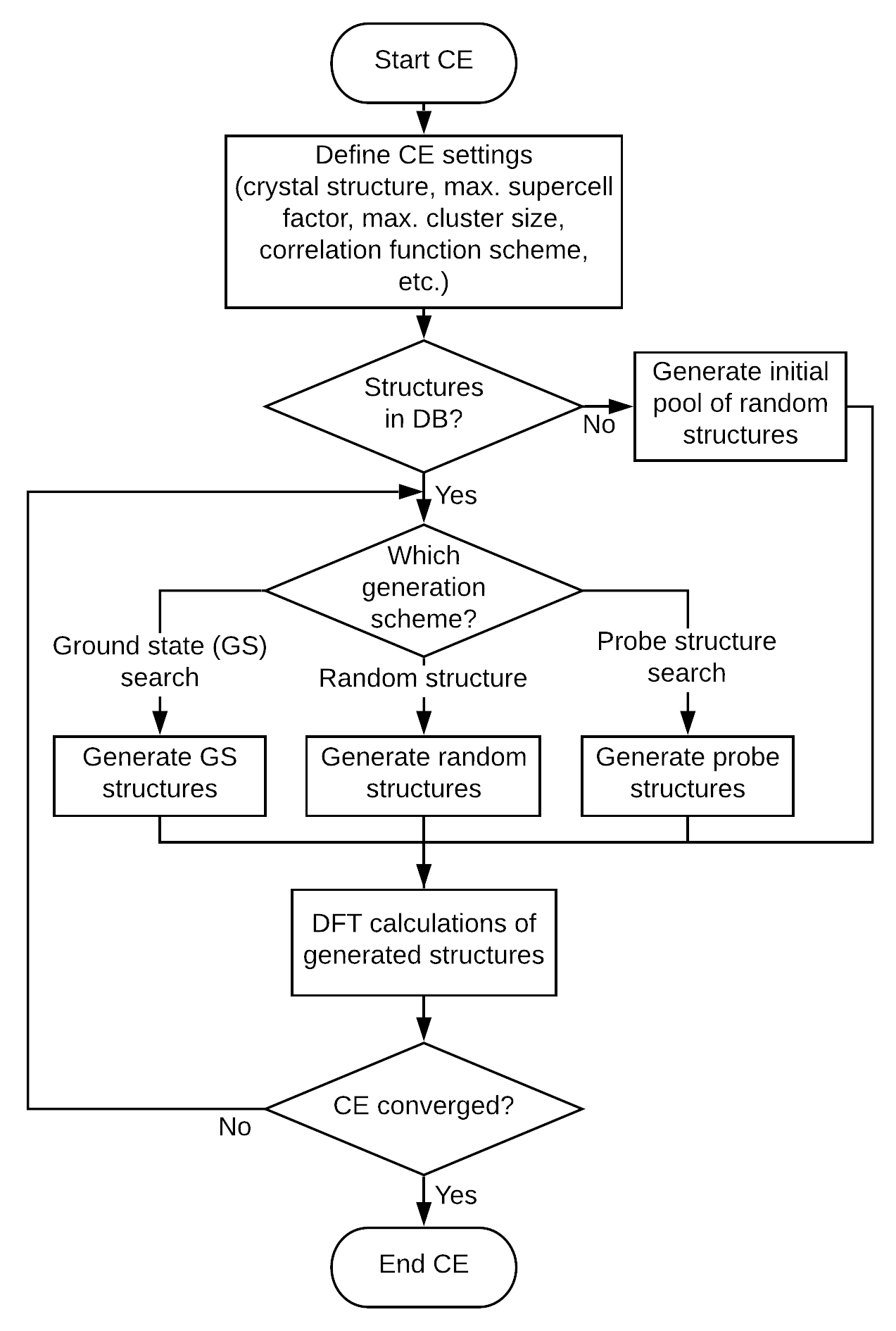}
	\caption{A flowchart of constructing and evaluating CE using CLEASE.}
\label{fig:ce_flowchart}
\end{figure}

A simple flowchart illustrating the procedure for constructing CE using CLEASE is shown in figure~\ref{fig:ce_flowchart}. The CLEASE workflow can be divided in to three main components: definition of CE settings, generation of training structures and evaluation of CE convergence. CLEASE takes an object-oriented approach where each component has its own class. The modular design approach not only enables easy implementation of new features but also makes the code flexible to use and intuitive to follow the CE construction and evaluation procedure shown in figure~\ref{fig:ce_flowchart}. A more detailed description of main components of the procedure is provided below.

\subsection{Definition of Cluster Expansion Settings}
The most fundamental component is to define which underlying crystal structure to use. ASE offers two functions to generate a crystal structure: \texttt{bulk} and \texttt{crystal}. The \texttt{bulk} function provides a simple way of generating common types of crystal structures by specifying the name of the crystal structure and its lattice constant value(s). The crystal structures supported by the \texttt{bulk} function are simple cubic, face-centered cubic, body-centered cubic, hexagonal close packed, diamond, zinc blende, rocksalt, cesium chloride, fluorite and wurtzite structures. For more complicated crystal structures, a \texttt{crystal} function is used to generate a crystal structure by providing its space group, lattice parameters and scaled coordinates of the unique atomic sites. The definitions of the cluster expansion settings are specified using \texttt{CEBulk} and \texttt{CECrystal} classes, which respectively calls \texttt{bulk} and \texttt{crystal} functions to generate an \texttt{Atoms} object with the user-specified crystal structure. 

The maximum size of the supercell (of the primitive cell) on which the DFT calculations are performed is also defined along with the definition of the underlying crystal structure. The maximum supercell size is specified using a \texttt{supercell\_factor} parameter, which is an integer corresponding to the product of the absolute values of expansion coefficients (integer weights of a general linear combinations of the unit cell vectors). In other words, if a unit cell has three vectors $\vec{a}$, $\vec{b}$ and $\vec{c}$, the configurations in the training set on which the DFT calculations are performed have cell vectors $\vec{a'}$, $\vec{b'}$ and $\vec{c'}$, which are defined as
\begin{align}
    \begin{split}
        \vec{a'} = i_1 \vec{a} + j_1 \vec{b} + k_1 \vec{c}\\
        \vec{b'} = i_2 \vec{a} + j_2 \vec{b} + k_2 \vec{c}\\
        \vec{c'} = i_3 \vec{a} + j_3 \vec{b} + k_3 \vec{c}
    \end{split}
    \label{eq:size_coefficients}
\end{align}
with integer coefficients $i_x$, $j_x$ and $k_x$, where $x\in\{1,2,3\}$. The \texttt{supercell\_factor} is then defined as
\begin{equation}
    \texttt{supercell\_factor} \geq \prod_{x=1}^{3} \prod_{y=1}^{3}\prod_{z=1}^{3}\left|i_x\right| \left|j_y\right| \left|k_z\right|,
    \label{eq:super_cell_condition}
\end{equation}
and all of the cells used in the training set should have coefficients satisfying the condition in (\ref{eq:super_cell_condition}). Only unique cell shapes are included in the pool by omitting the cells that can be mapped on to the existing cells in the pool by rotation and reflection. The use of supercells with varying sizes and shapes enables the exploration of a larger configurational space without adding extra computational burden compared to using one fixed supercell size and shape. A set of training structures for CE are later generated iteratively from the pool of possible structures that are realizable using these supercells. To reduce the required computational resources, the structures using smaller supercells are generated (and calculated) first, followed by the larger supercells. The users also have a flexibility to select the supercell size using an optional \texttt{size} parameter, which is a $3 \times 3$ matrix (or a nested list in Python) specifying the values of the integer coefficients in (\ref{eq:size_coefficients}). 

Theoretically, an infinite number of clusters can be generated for a given system. The number of clusters is limited to a finite size in practice, and CLEASE takes an approach to generate all possible clusters that are under the truncation threshold (i.e., a maximum number of atoms in clusters and maximum diameter) specified by the user. A whole or subset of the generated clusters is selected during the convergence evaluation process. By default, up to four-body clusters (i.e., empty, one-, two-, three- and four-body clusters) with their diameters up to 5 \AA{} are generated. The users have an option to define their own threshold settings both at the beginning of the CE procedure and at a later stage of the CE iteration cycles. CLEASE also offers a feature to visualize the generated clusters in order to assist the user to develop an intuition on the generated clusters. 

Within the CE formalism, there does not exist a unique set of definitions for basis functions; the basis functions are considered valid if they form a complete set. Consequently, several definitions are used in practice. The two most widely used definitions are the original definitions by Sanchez et al. \cite{Sanchez1984} and the one later developed by van de Walle \cite{VandeWalle2009}, which is used in the Alloy Theoretic Automated Toolkit (ATAT) \cite{VandeWalle2009, VandeWalle2002a}. The two definitions are equally valid, and both are implemented in CLEASE. 

CLEASE offers an option to ignore a set of symmetrically inequivalent atomic sites if they are always occupied by one element type for all possible configurations. The contributions of these atoms are not explicitly included in the cluster expansion and are automatically included in the constant term ($J_0$) in (\ref{eq:cluster_expansion2}). For example, lithium metal oxides (\ce{LiMO2}) with first-row transition metals (M = \{Sc, Ti, V, Cr, Mn, Fe, Co, Ni, Cu\}) have a rocksalt lattice structure \cite{Urban2016, Urban2014} with an exception of \ce{LiMnO2}, which is orthorhombic \cite{Hewston1987}. The rocksalt lattice structure consists of two face-centered cubic sublattices. For the case of the cation-disordered rocksalt lattice \ce{LiMO2}, one sublattice is occupied by lithium and other metal atoms while the other is occupied by oxygen atoms. The complexity of the CE model of such systems can be reduced to a cation sublattice consisting only of two element types (the oxygen sublattice is ignored). As such, an optional Boolean argument is present in CLEASE to enable/disable the reduction of the complexity of the model by ignoring the such atoms if they exist in the system.  

A range of compositions (or concentration) of the system to be studied is specified using a \texttt{Concentration} class. First, the constituting elements of the system are categorized into the basis which they belong. For example, \ce{LiVO2} in a rocksalt lattice structure is expressed using two lists: [`\ce{Li}', `\ce{V}'] and [`\ce{O}']. It is noted that CE needs to keep track of the location of vacancies when they are present in the system. The location of vacancies are tracked by treating a vacancy as a regular atom with its atomic symbol set to `X' or atomic number set to zero. The \ce{LiVO2} with \ce{Li} vacancies is then expressed using [`\ce{Li}', `\ce{V}', `X'] and [`\ce{O}']. 

The range of each element (including vacancies) can be specified in one of the two convenient methods built in to the \texttt{Concentration} class. The simplest method is to specify the concentration range of each constituting element by calling \texttt{set\_conc\_ranges} method in \texttt{Concentration} class. For the cases where concentrations of two or more elements depend on one another, one can specify concentration range using \texttt{set\_conc\_formula\_unit} method where the relationships between the concentrations of two or more elements can be expressed in a list of strings. For the example of \ce{LiVO2} with \ce{Li} vacancies, a list of strings that specifies relationship between the number of \ce{Li} atoms and the number of vacancies,  [``Li$<$x$>$V$<$1$>$X$<$1-x$>$'', ``O$<$2$>$''], is passed as an argument to the \texttt{set\_conc\_formula\_unit} method. Another argument specifying the range of the concentration variable, e.g., \{`x': (0, 1)\}, is also passed to the \texttt{set\_conc\_formula\_unit} method in order to specify the concentration range of Li and Li vacancies. The concentration ranges specified by either \texttt{set\_conc\_ranges} or \texttt{set\_conc\_formula\_unit} methods are internally interpreted in the \texttt{Concentration} class as a list of linear equations that specify (1) the relationships of the concentrations of constituting elements and (2) their upper/lower bounds. The advanced users can alternatively specify the coefficients of the linear equations used in the \texttt{Concentration} class if a greater flexibility is needed in specifying the concentration ranges.

\subsection{Generation of Training Structures}
CLEASE uses \texttt{NewStructures} class to generate training structures, which provides three different methods perform the task. The first and most trivial method is to generate a set of random structures. This method serves to generate an initial pool consisting of a small number of structures. The random generation method is used in the first iteration cycle of CE construction as shown in figure~\ref{fig:ce_flowchart}. An initial cluster expansion is capable of making a first set of predictions albeit with a low accuracy. It is noted that all of the generated training structures, along with their correlation function values, are stored in a database file. 

Once the initial CE is constructed, the user is given three different choices for introducing an additional set of training structures. The first and most straightforward option is to keep generating random structures. Although trivial, generating random structures is claimed to be the best strategy when compressive sensing is used to select clusters \cite{Nelson2013a}. The second method is to generate ground-state and other low-energy structures based on current cluster expansion (i.e., based on the pool of structures already calculated) \cite{van2002automating}, which have the enthalpies of formation either on or near the convex hull \cite{Urban2016a}. The inclusion of ground-state and near-ground-state structures serves an important purpose of increasing the accuracy in predicting the correct ground states. A global minimization technique can be used to generate (near) ground-state configurations, and CLEASE uses a simulated annealing technique.

The last method of generating the training set is referred to as a ``probe structure'' method \cite{Seko2009a, Seko2014}. The probe structure method introduces a new structure that minimizes the mean variance of the predicted physical quantity $q(\bm{\sigma})$. The mean variance of the predicted quantity $q$ is written as \cite{Seko2009a}
\begin{align}
    \begin{split}
        \mathrm{Var}[\hat{q}_i] &= \frac{1}{N_\mathrm{config}} \sum_{i=1}^{N_\mathrm{config}} [\mathbf{X}_i (\mathbf{X}^T\mathbf{X})^{-1} \mathbf{X}_i^T]e^2\\
        &= \{\mathrm{tr}[(\mathbf{X}^T\mathbf{X})^{-1} \Sigma ] + \mu (\mathbf{X}^T\mathbf{X})^{-1} \mu^T \} e^2 \\
        &= \Lambda \cdot e^2,
    \end{split}
\end{align}
where $e^2$ is the variance of the error in the training set, $\Sigma$ is the covariance matrix of the correlation functions of the training set and $\mu$ is a vector of the mean correlation functions of the structures in the training set. The probe structure is the one that reduce the value of $\Lambda$ the most when introduced to the training set, which is found using the simulated annealing procedure. 

The newly generated structures are compared with the existing structures in the training set in order to avoid introducing duplicate structures. We adopted the structure comparison algorithm developed by Lonie and Zurek \cite{Lonie2012} to identify equivalent structures. It is desirable to have the new structure compared against the existing structures in the training set as efficiently as possible. As a first step, the structures that have different chemical composition than the newly generated structure are filtered, and the new structure is compared only with the remaining structures. Once the candidate transformations for mapping the new structure onto one structure in the database are identified using the algorithm suggested by Lonie and Zurek, we note that exactly the same transformations can be used for the remaining structures in the database. Therefore, the structure comparison algorithm implemented in ASE is optimized for the case where one structure is to be compared against many.
 
In addition to the aforementioned methods of generating the training structures, CLEASE also offers a built-in function to import structures to the database. The import function also has an option to specify the calculated $q$ value, which allows users to easily import the previously calculated results.

\subsection{Evaluation of Cluster Expansion Convergence}

An evaluation process to determine the convergence of CE includes a selection of clusters, a determination of their ECI values and an assessment of the LOO or $k$-fold CV score using the selected clusters and their ECI values. An entire evaluation process is performed using an \texttt{Evaluate} class. 

The simplest way to determine the ECI values of the generated clusters is by using OLS to minimize residual sum of squared errors (RSS). It is highly likely that the ECI values found using OLS are overfitted. Therefore, Bayesian compressive sensing and $\ell_1$ and $\ell_2$ regularization methods are implemented, and it is highly recommended to use a regularization methods to select clusters and evaluate their ECI values. 

A default option in the \texttt{Evaluate} class is to include all of the clusters generated using the cluster truncation conditions specified in \texttt{CEBulk} or \texttt{CECrystal} class, and either the entire or a subset of these clusters are selected for fitting  depending on the method used. The \texttt{Evaluate} class provides additional options in which the users can select a subset of the generated clusters to perform any of OLS, Baysian compressive sensing and $\ell_1$ and $\ell_2$ regularization. The first option is by manually specifying which clusters to include, while the second option is to provide a stricter truncation conditions than the ones set in the \texttt{CEBulk} or \texttt{CECrystal} class. The first option allows the \texttt{Evaluate} class to be used in conjunction with other cluster selection methods such as genetic algorithm. For example, a user can optionally use genetic algorithm (included in CLEASE as a separate \texttt{GAFit} class) to pre-screen a large cluster pool and subsequently pass a subset of clusters to the \texttt{Evaluate} class. The feature to freely select a subset of a large pool of clusters along with the use of OLS, Bayesian compressive sensing and $\ell_1$ and $\ell_2$ regularization methods allows the users to easily experiment with various settings to understand how the system behaves and to optimize the ECI values for achieving the lowest LOOCV score.

To further assist the evaluation process, the \texttt{Evaluate} class contains two built-in methods that automatically determine the LOOCV when a regularization method is used. The first method, \texttt{plot\_fit}, determines and stores the selected clusters and their ECI values for a value of regularization parameter ($\lambda$) specified by the user. It also plots the fit of all data points in the training set to their calculated values and presents the LOO/$k$-fold CV score of the specified $\lambda$ value. Since the most cumbersome task in determining the convergence of CE is finding the optimal $\lambda$ value that yields the lowest CV score, another method, \texttt{plot\_CV}, is also implemented. It takes a range and number of $\lambda$ values to evaluate as inputs and returns the best $\lambda$ value in the specified range along with its LOO/$k$-fold CV score. The \texttt{plot\_CV} method also plots LOO/$k$-fold CV score as a function of $\lambda$ and provides an option to store the results in a log file such that the users can add more $\lambda$ values to the list at a later stage without having to re-evaluate the same $\lambda$ values in the process.  

\subsection{Metropolis Monte Carlo and Simulated Annealing}
The user can perform statistical sampling of the system on a larger simulation cell once the cluster expansion is constructed. The final selection of cluster and their ECI values can be stored and passed to other classes to conduct statistical analyses. A separate \texttt{Calculator} class for cluster expansion is implemented in ASE. The \texttt{Clease} calculator class takes a list of clusters and their ECI values as inputs, and the users can select what type of trial moves are allowed. The sampling in the canonical ensemble allows the swapping two atoms with different constraint conditions (i.e., swap any two atoms, swap any two atoms in the same basis, swap two nearest neighbors, swap two nearest neighbors in the same basis) while the semi-grand canonical ensemble allows changing the type of occupying element at a random site. 

The evaluation of the physical quantity $q(\bm{\sigma})$ is performed using (\ref{eq:cluster_expansion3}), which is a fast because the \texttt{Clease} calculator keeps track of the changes in the \texttt{Atoms} object to update the correlation functions. When the physical quantity being modeled is energy, a trial move of the standard Metropolis algorithm has an acceptance probability \cite{thijssen2007computational}
\begin{equation}
    P_\mathrm{acc} = \mathrm{min}\left\{1, \exp\left(\frac{-(E_\mathrm{new} - E_\mathrm{old})}{k_BT}\right)\right\},
    \label{eq:accept}
\end{equation}
where $E_\mathrm{new}$ and $E_\mathrm{old}$ are the energy of the new and old configuration, respectively. As the \texttt{Clease} calculator keeps track of the change in the \texttt{Atoms} object after each move, updating the correlation functions is restricted to the contributions of one and two atoms for the semi-grand canonical ensemble and canonical ensemble, respectively. 

\section{Examples}
\label{sec:example}
Here, we present two example systems to illustrate the capabilities of the CLEASE code. The first example illustrates the investigation of a Au--Cu binary alloy. The second example shows the cluster expansion on a more complex \ce{Li2CrO2F} system consisting of four types of elements and vacancy. All of the interactions of cluster expansions are computed from DFT calculations of energies, and the computational settings used for generating the results shown in this section are specified in section~\ref{sec:methods}.

\subsection{Au--Cu Alloy}
\begin{figure}[!htb]
	\centering
	\includegraphics[width=86mm]{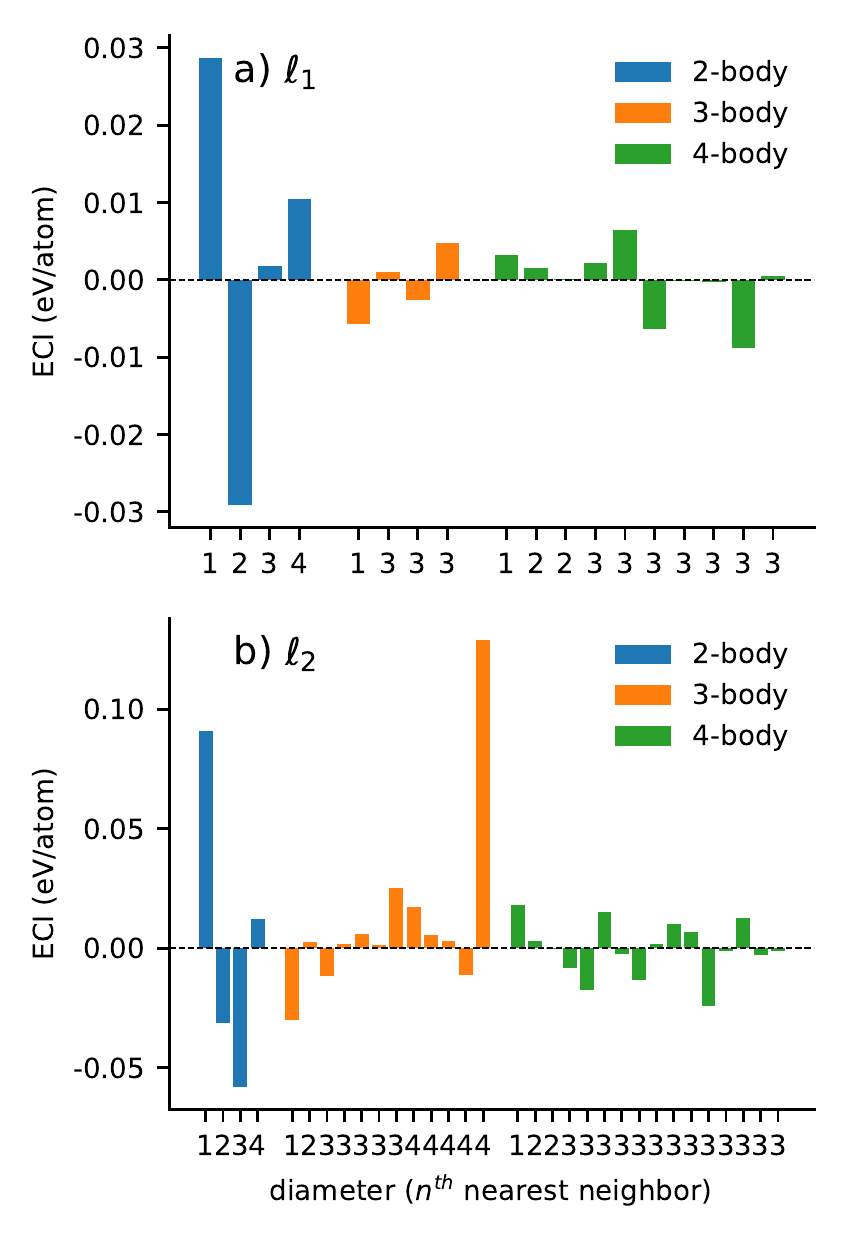}
	\caption{ECIs obtained via a) $\ell_1$ regularization and b) $\ell_2$ regularization.}
\label{fig:eci_aucu}
\end{figure}

The binary Au--Cu alloy system is studied at temperatures ranging from 100 K to 800 K over the entire composition range. The resulting values obtained for both $\ell_1$ and $\ell_2$ regularization are shown in figure~\ref{fig:eci_aucu}. The ECI value of the empty cluster is found to be $-3.49$ eV/atom for both cases, and the ECI value of the one-body cluster is 0.27 eV/atom and 0.13 eV/atom for $\ell_1$ and $\ell_2$ regularization, respectively. The ECI values of empty and one-body clusters are not included in figure~\ref{fig:eci_aucu} for better visibility. The LOOCV score for the $\ell_1$- and $\ell_2$-regularized fit were 4.49 meV/atom and 4.67 meV/atom, respectively. The $\ell_1$ regularization scheme yields a slightly lower CV score despite having a smaller number of clusters ($\ell_1$-regularized fit has 20 clusters while the $\ell_2$-regularized fit has 34 clusters). 

A qualitative information on the thermodynamic behavior of the system can be extracted by inspecting the ECI values for simple binary system. Based on the fact that the energetically favorable configurations have DFT energies that are more negative than less favorable ones and that the two site variables are $+1$ or $-1$, one can infer that a positive ECI value of the pair interaction term means that a pair consisting of two different elements is energetically preferred at a low temperature. It can be seen in figure~\ref{fig:eci_aucu} that the ECIs of the nearest-neighbor and second nearest-neighbor pairs are positive and positive, respectively. The signs indicate that the Au--Cu system energetically favors the strong mixing of the constituting elements such that the alternating patterns found in L1$_0$- and L1$_2$-type ordered structures are likely to emerge, which is in a good agreement with experimental and computational observations \cite{van2002automating, Wei1987, Ozolins1998_1, Ozolins1998_2, Wolverton1998, Lysgaard2015, massalski1986binary, Hultgren1973, Fedorov2016}.

\begin{figure}[!htb]
	\centering
	\includegraphics[width=86mm]{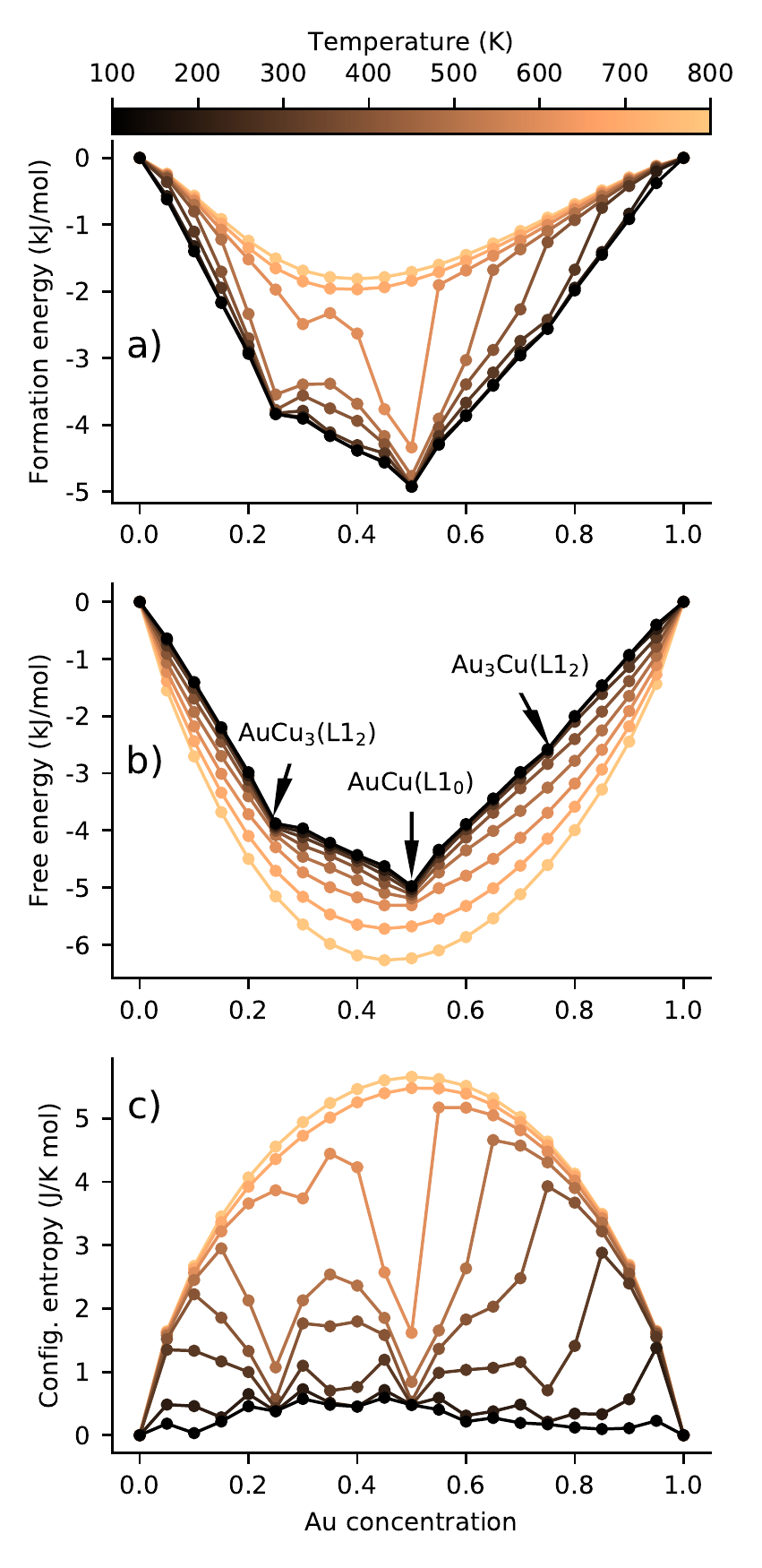}
	\caption{Thermodynamic quantities for the Au--Cu computed for temperatures ranging from 100 K to 800 K over the entire composition range. a) Formation energy. b) Free energy of formation. c) Entropy.}
\label{fig:aucu_thermo}
\end{figure}

It is experimentally determined that Au--Cu alloys have three ordered phases at low temperatures \cite{massalski1986binary, Hultgren1973, Fedorov2016}: \ce{AuCu3}, \ce{AuCu} and \ce{Au3Cu}. Furthermore, the transition temperatures for \ce{AuCu3}, \ce{AuCu} and \ce{Au3Cu} are reported to be 663 K, 683 K and $\sim$490 K, respectively, and they are often used as reference values for assessing the computational models \cite{van2002automating, Wei1987, Ozolins1998_1}.  The formation energy, free energy of formation and configurational entropy are obtained through Metropolis Monte Carlo simulations and are shown in figure~\ref{fig:aucu_thermo}. As the CE is trained with fully relaxed structures (zero pressure), the formation energy is determined using 
\begin{equation}
    \Delta U = U - xU_{\mathrm{Au}} - (1-x)U_{\mathrm{Cu}},
\end{equation}
where $U$ is the internal energy of the configuration, $x$ is the gold concentration, $U_{\mathrm{Au}}$ is the internal energy of pure gold and $U_{\mathrm{Cu}}$ is the internal energy of pure copper. Similarly, the free energy of formation is obtained by subtracting the weighted average of the free energy for the pure phases. The configurational entropy is given by the difference between the internal energy and the free energy, divided by the temperature at which the Monte Carlo is sampled. The three ordered phases (\ce{AuCu3}, \ce{AuCu} and \ce{Au3Cu}) are found on the convex hull of the free energy of formation in figure~\ref{fig:aucu_thermo}b. Furthermore, the entropy of the ordered phases form local minima as shown in figure~\ref{fig:aucu_thermo}c. As the temperature increases, the free energy becomes a smooth convex curve with a minimum at around $50\%$ composition, and the system is in a random phase with no short-range order.

\begin{figure}[!htb]
	\centering
	\includegraphics[width=86mm]{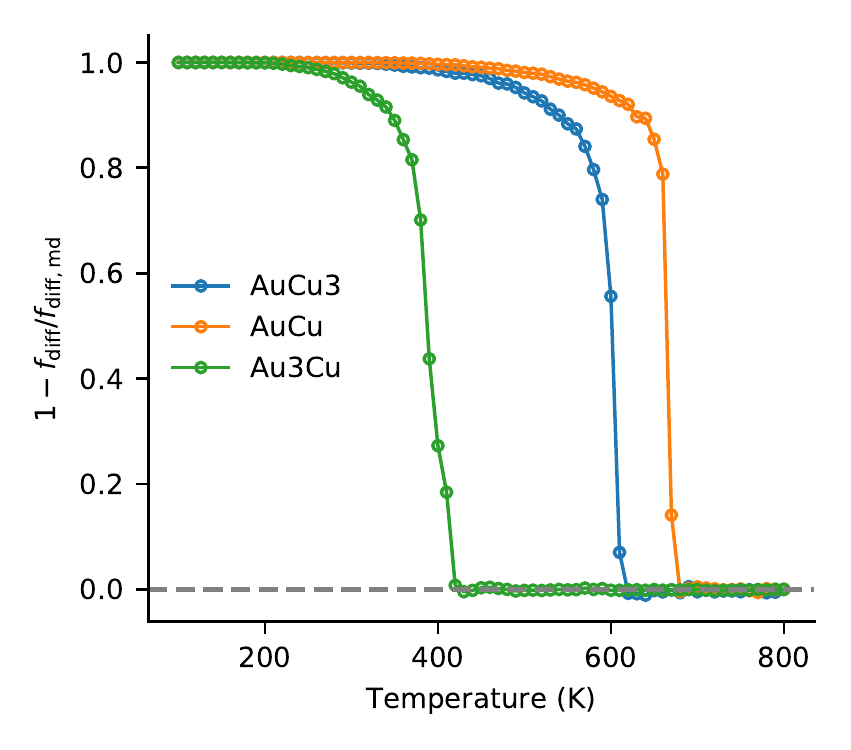}
	\caption{Order parameter as a function of temperature (1 in an ordered phase and 0 in a random phase).}
\label{fig:order_param}
\end{figure}

An accurate estimate of the order/disorder transition temperature can be found by tracking the evolution of an order parameter. The average fraction of sites having a different element than the same site in the ground state, $f_\mathrm{diff}$, is tracked as the system evolves. $f_\mathrm{diff}$ is normalized by the expected fraction of different sites in a random phase, $f_\mathrm{diff, rnd}$, and an order parameter, $\eta$, is defined as
\begin{equation}
    \eta = 1 - f_\mathrm{diff}/f_\mathrm{diff, rnd}.
\end{equation}
The order parameter is is used for detecting the phase transition as shown in figure~\ref{fig:order_param}. The computationally predicted order/disorder transition temperature of \ce{AuCu3}, \ce{AuCu} and \ce{Au3Cu} are around 600 K, 665 K and 385 K, respectively, which are in a good agreement with the experimental reference values \cite{massalski1986binary, Hultgren1973, Fedorov2016}. 

\begin{figure}[!htb]
	\centering
    \includegraphics[width=86mm]{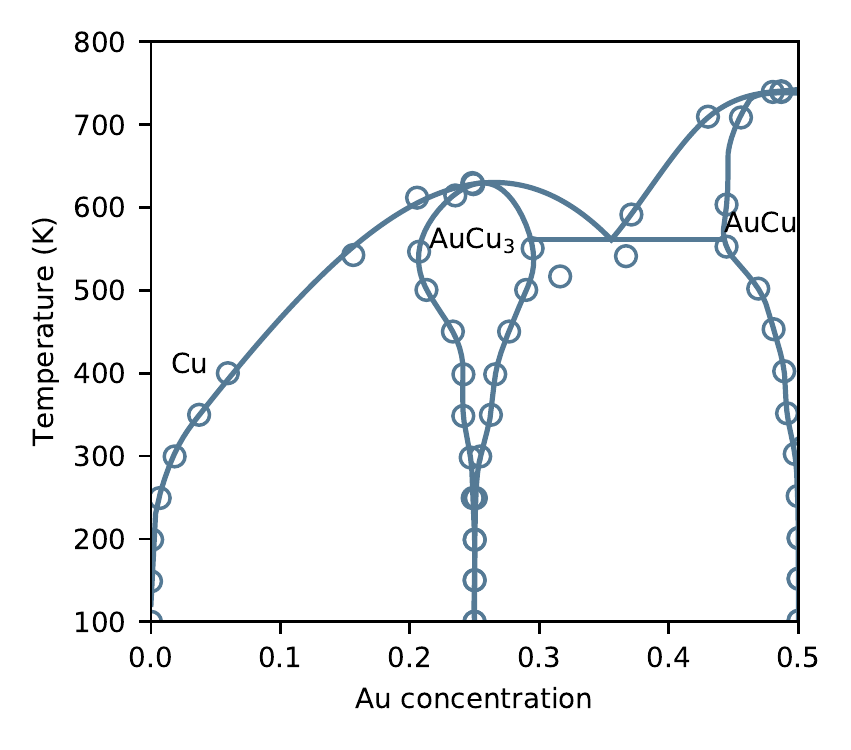}
    \caption{Phase diagram of Au$_x$Cu$_{1-x}$ where $0 \leq x \leq 0.5$. Circles are computed phase boundary points and lines are spline fits of the computed boundary points.}
\label{fig:phase_diagram}
\end{figure}

One of the most common way to describe the characteristics of a binary alloy is by constructing a phase diagram. A phase diagram can be generated computationally using a semi-grand canonical MC where a grand potential is obtained via thermodynamic integration in (\ref{eq:G_int_C}) at fixed chemical potentials. The integration starts from the low temperature limit for the ordered phases and from the high temperature limit for disordered phases where the free energy per atom is given by $k_BT \ln 2$. The phase boundary between two phases is identified by locating the intersection point between the grand potential in the two co-existing phases. The phase diagram generated via semi-grand canonical MC is shown in figure~\ref{fig:phase_diagram}. The phase diagram closely resembles the phase diagrams constructed from the experimental measurements \cite{massalski1986binary, Hultgren1973, Fedorov2016} and is also in a qualitatively agreement with the phase diagrams constructed from computational results \cite{van2002automating, Wei1987, Walle2002a}.

\subsection{Lithium Chromium Oxyfluoride}
One of the recent focus areas of lithium-ion battery research is the development of high-capacity cathode materials. Lithium metal oxyfluorides (\ce{Li2MO2F}, $\mathrm{M} = $\{V, Cr, Mn, Ti, Ni, $\ldots$\}) is a family of materials that is at the forefront of the current research. The challenges for studying \ce{Li2MO2F} is in the vast size of the configurational space, which exhibit not only the cation disorder commonly found in lithium metal oxides \cite{Abdellahi2016, Urban2016} but also anion disorder which is also present due to the mixed O/F composition \cite{Chen2015, Chen2016}. The fact that the underlying crystal structure of \ce{Li2MO2F} can vary at different lithiation levels \cite{Cambaz2016} adds the complexity to investigate their properties. It is, however, known that the most predominant crystal structure is of disordered rocksalt type \cite{Ren2015}, particularly at high-lithiation levels. We therefore show an example CE study of \ce{Li2CrO2F} in a rocksalt lattice configuration.

\begin{figure}[!htb]
	\centering
    \includegraphics[width=73mm]{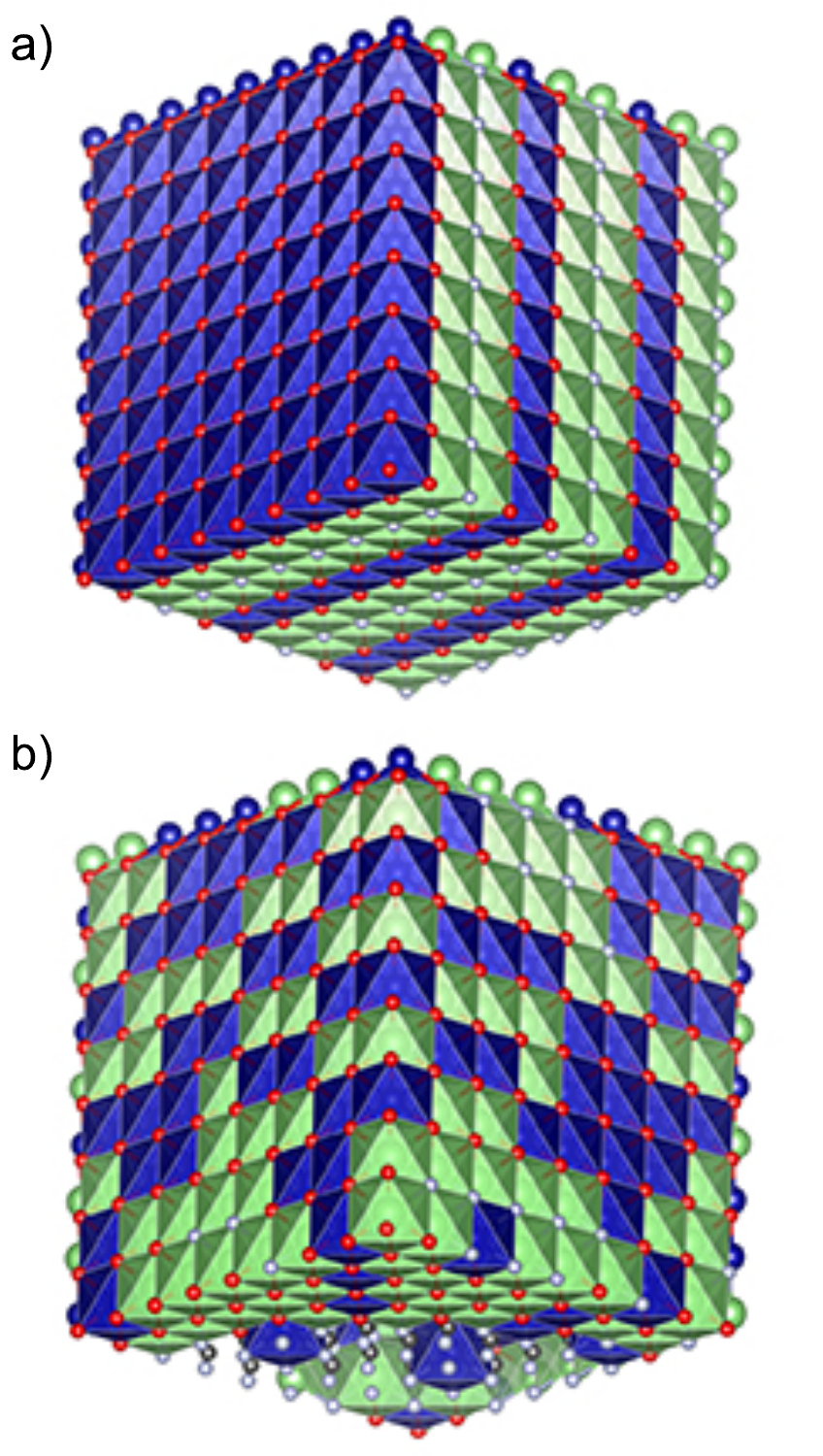}
    \caption{A snapshot of Li$_{x}$CrO$_2$F during the Monte Carlo run at 293 K where $x$ is a) 2.0, b) 1.5. The Li atoms are shown in green, the Cr atoms are shown in blue, the oxygen atoms are shown in red and the F atoms are shown in white.}
\label{fig:licro2f_structures}
\end{figure}

The Monte Carlo annealing study reveals that \ce{Li2CrO2F} (i.e., fully lithiated compound) takes a layered structure at room temperature (293 K) as shown in figure~\ref{fig:licro2f_structures}a. The layer structure shows a  $\ldots$--Li--F--Li--O--Cr--O--$\ldots$ pattern, which is similar to a $\ldots$--Li--O--M--O--$\ldots$ layered pattern observed in lithium metal oxides \cite{Urban2016, Mizushima1981, VanDerVen2004}. The layered structure is lost upon delithiation, which leads to disordered structures as shown in figure~\ref{fig:licro2f_structures}b. The emergence of disordered structures agrees well with the previous experimental observations \cite{Chen2016, Ren2015}, and it is important to model the disordered atomic arrangement as it has a direct link to the Li transport mechanism (e.g., a presence of zero-transition-metal pathways \cite{Urban2016, Urban2014, Lee2014}). 

Thermodynamics quantities of Li$_x$CrO$_2$F can be extracted with the same procedure described for the Au--Cu system. One of the most crucial thermodynamic parameters for characterizing cathode materials for Li-ion batteries is the free energy as it is directly linked to the operating voltage of the cell. The operating voltage of Li$_x$CrO$_2$F is defined as
\begin{align}
    \begin{split}
        \mathrm{voltage} &= - \frac{ \mu_\mathrm{Li}^\mathrm{cathode} - \mu_\mathrm{Li}^\mathrm{anode}}{e}\\
        &= - \frac{\frac{\mathrm{d}G_{\mathrm{Li}_x\mathrm{Cr}_2\mathrm{F} }}{\mathrm{d}x} - \mu_\mathrm{Li}^\mathrm{anode}}{e},
    \end{split}
    \label{eq:voltage}
\end{align}
where $\mu_\mathrm{Li}$ is the chemical potential in eV per Li atom, $e$ is an electron charge and $G_{\mathrm{Li}_x\mathrm{Cr}_2\mathrm{F}}$ is the free energy of Li$_x$CrO$_2$F in eV per formula unit. Li metal is used as an anode and thus, $\mu_\mathrm{Li}^\mathrm{anode}$ is constant. 

\begin{figure}[!htb]
	\centering
    \includegraphics[width=86mm]{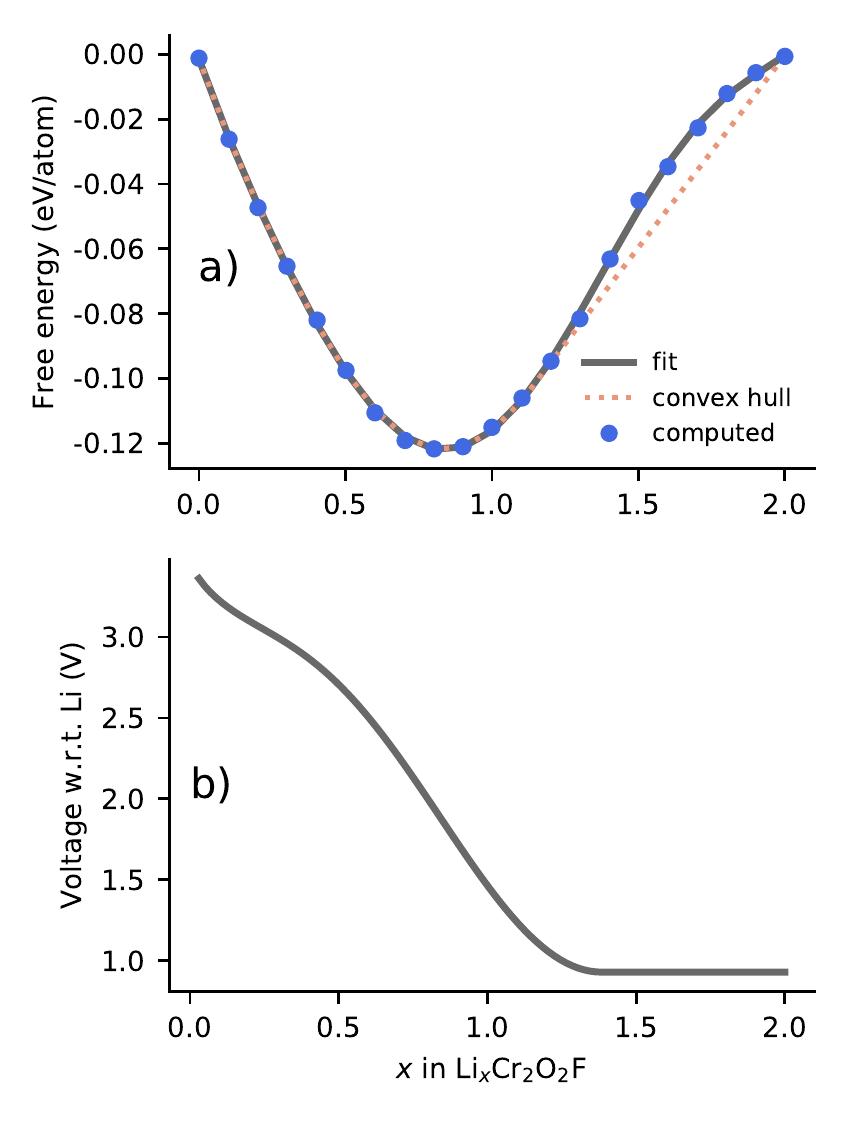}
    \caption{Free energy of Li$_{x}$CrO$_2$F and its voltage with respect to Li metal at 293 K.}
\label{fig:free_energy_voltage}
\end{figure}

The free energy of Li$_x$CrO$_2$F and its voltage profile at 293 K are shown in figure~\ref{fig:free_energy_voltage}. The free energy in figure~\ref{fig:free_energy_voltage}a has three parts: free energy values computed from MC simulations, a smooth curve fitted to the computed values (using Redlich-Kister polynomials \cite{RK_fit}) and a convex hull of the fitted curve. The curve fit is used for generating the voltage plot because the derivative of the free energy used for calculating the voltage values are susceptible to small noise that are present in the MC simulation results. Furthermore, a range in which the free energy curve is above the convex hull represents the region where a phase transition occurs: the cathode forms a mixture of two phases at which the fitted curve and the convex hull intersect. The voltage profile in figure~\ref{fig:free_energy_voltage}b is generating using (\ref{eq:voltage}) where the values on the convex hull are used for $G_{\mathrm{Li}_x\mathrm{Cr}_2\mathrm{F}}$. The voltage profile in figure~\ref{fig:free_energy_voltage}b is in a good agreement with those observed experimentally \cite{Chen2016, Ren2015}.

\section{Methods}
\label{sec:methods}

\subsection{Density Functional Theory Calculations}
All of the calculations are performed with the Vienna Ab initio Simulation Package (VASP) \cite{VASP1, VASP2, VASP3, VASP4} using the projector augmented-wave (PAW) method \cite{Blochl1994a}. The generalized gradient approximation as parametrized by Perdew, Burke and Ernzerhof \cite{Perdew1996} is used as the exchange-correlation functional. It is important to have a consistent and accurate dataset (i.e., DFT calculations with high energy cutoff and $k$-point mesh density) in order to minimize the numerical noise introduced to the CE training. The plane-wave cutoff of 500 eV is used, and both the cell and atomic positions are fully relaxed such that all the forces are smaller than 0.02 eV/\AA{}. A rotationally invariant Hubbard \textit{U} correction \cite{Anisimov1991, Cococcioni2005a} is applied to the $d$ orbital of Cr with the $U$ value of 3.7 eV. The calculations are performed with supercells containing up to 18 and 54 atoms for Au--Cu alloy and \ce{Li2CrO2F} systems, respectively. Integrations over the Brillouin zone were carried out using the Monkhorst-Pack scheme \cite{Monkhorst1976} with a grid with a maximal interval of 0.04 \AA{}$^{-1}$.

\subsection{Cluster Expansion Model}
The CE model for Au--Cu alloy and Li$_x$CrO$_2$F are trained using 34 and 390 DFT calculations, respectively. CE model is trained for the entire composition range of Au--Cu alloy (from pure Au to pure Cu) and Li$_x$CrO$_2$F on a rocksalt lattice with $x$ ranges from 0 to 2. Up to four-body clusters with the maximum diameter of 6.0 \AA{} are generated for Au--Cu alloy. Up to four-body clusters are generated for Li$_x$CrO$_2$F with the maximum diameter of 7.0 for two- and three-body clusters and 4.5 \AA{} for four-body clusters. $\ell_1$ and $\ell_2$ regularization schemes with the regularization parameter ranging from $10^{-7}$ to $10^2$ are assessed at various maximum radii to find the optimal setting that leads to the lowest LOOCV score. For the Au--Cu alloy, $\ell_1$ regularization with the maximum diameter of 6.0 \AA{}, 5.0 \AA{} and 5.0 \AA{} for 2-, 3- and 4-body clusters, respectively, yields the lowest LOOCV score of 4.49 meV/atom. The minimum LOOCV score achieved using $\ell_2$ regularization scheme is 4.67 meV/atom when the maximum diameter is set to 6.0 \AA{}, 6.0 \AA{} and 5.0 \AA{} for 2-, 3- and 4-body clusters, respectively. Similarly, $\ell_1$ regularization performed better than $\ell_2$ regularization on Li$_x$CrO$_2$F with the lowest LOOCV score of 21.38 meV/atom (maximum diameter set to 7.0 \AA{}, 7.0 \AA{} and 4.5 \AA{} for 2-, 3- and 4-body clusters, respectively). It is noted that although the LOOCV of Li$_x$CrO$_2$F seems larger compared to that of Au--Cu, it should be taken into account that the cohesive energy of metallic alloys are in general much smaller than those of oxyfluorides.

\subsection{Metropolis Monte Carlo Simulations}

For Au--Cu alloy, Metropolis Monte Carlo simulations are carried out using a $10 \times 10 \times 10$ supercell consisting of 1,000 atoms for determining thermodynamic quantities. The system is equilibrated with 100 sweeps, and an average energy is collected through an additional 2,000 sweeps at each temperature for determining the thermodynamic quantities. A $30 \times 30 \times 30$ supercell consisting of 27,000 atoms is used to determine the transition temperatures and to construct a phase diagram. The transition temperatures are determined by equilibrating the systems with 100 sweeps, followed by sampling the order parameter via an additional 1,000 sweeps. A phase diagram is generated by performing a semi-grand canonical MC, where the system is equilibrated using 100 sweeps, followed an additional 1,000 sweeps to obtain an average semi-grand canonical energy at each temperature at a fixed chemical potential. A $9 \times 9 \times 9$ cell consisting of 1,458 atoms is used for Li$_x$CrO$_2$F. The temperature is gradually lowered from 10,000 K, and the structures are equilibrated at each temperature via 100 sweeps to ensure that the system is equilibrated before sampling. The average energy is then sampled via 1,000 sweeps at each temperature. 

\section{Conclusions}
We present the implementation of CLEASE, which fully integrates the cluster expansion method to ASE package. The aim of the developed code is to make cluster expansion more accessible to non-specialists and to incorporate modern machine learning techniques to cluster expansion method in one comprehensive and versatile package. The use of the popular Python programming language and implementing the code as a part of widely used ASE package lowers the barrier for the newcomers to the field to easily learn and use CE as a part of their research methods. By automatically generating clusters and calculating the correlation functions of both semi-automatically generated and user-supplied structures, it minimizes both the possible introduction of user errors and complicated process of constructing/evaluating the cluster expansion. The capability of CLEASE is presented with two example usage cases with a different level of system complexity. The examples demonstrate that CE can correctly predict the material behavior that require statistical sampling on a large simulation cell. 

\section{Acknowledgements}
This project has received funding from the European Union's Horizon 2020 research and innovation programme under grant agreement No 711792 (FET‐OPEN project LiRichFCC). Authors thank valuable discussions with Dr. Juhani Teeriniemi and Prof. Kari Laasonen. JMGL acknowledges support from the Villum Foundation's Young Investigator Programme (4$^{th}$ round, project: \textit{In silico design of efficient materials for next generation batteries.} Grant number: 10096).

\end{document}